\documentclass{article}
\usepackage{arxiv}
\usepackage[utf8]{inputenc} 
\usepackage[T1]{fontenc}    
\usepackage{hyperref}       
\usepackage{url}            
\usepackage{booktabs}       
\usepackage{amsfonts}       
\usepackage{nicefrac}       
\usepackage{microtype}      
\usepackage{lipsum}

\usepackage{cite}
\usepackage[margin=0.5cm]{caption}
\usepackage{graphics}
\usepackage{graphicx}
\usepackage{subcaption}
\usepackage{tikz}
\usepackage{float}
\usepackage{amsmath}
\usepackage{amssymb}
\usepackage{verbatim}
\usepackage{multirow}

\title{PDCOVIDNet: A Parallel-Dilated Convolutional Neural Network Architecture for Detecting COVID-19 from Chest X-Ray Images}

\author{Nihad Karim Chowdhury\\ Department of Computer Science and Engineering\\ University of Chittagong\\ Bangladesh\\ \texttt{nihad@cu.ac.bd}\\
   \And
 Md. Muhtadir Rahman\\ Department of Computer Science and Engineering\\University of Chittagong\\Bangladesh\\\texttt{muhtadirrahman@gmail.com}\\
  \And
 Muhammad Ashad Kabir\\School of Computing and Mathematics\\Charles Sturt University,NSW\\Australia\\\texttt{akabir@csu.edu.au}\\
}

\begin{document}
\maketitle

\begin{abstract}
The COVID-19 pandemic continues to severely undermine the prosperity of the global health system. To combat this pandemic, effective screening techniques for infected patients are indispensable. There is no doubt that the use of chest X-ray images for radiological assessment is one of the essential screening techniques. Some of the early studies revealed that the patient's chest X-ray images showed abnormalities, which is natural for patients infected with COVID-19. In this paper, we proposed a parallel-dilated convolutional neural network (CNN) based COVID-19 detection system from chest x-ray images, named as Parallel-Dilated COVIDNet (PDCOVIDNet). First, the publicly available chest X-ray collection fully preloaded and enhanced, and then classified by the proposed method. Differing convolution dilation rate in a parallel form demonstrates the proof-of-principle for using PDCOVIDNet to extract radiological features for COVID-19 detection. Accordingly, we have assisted our method with two visualization methods, which are specifically designed to increase understanding of the key components associated with COVID-19 infection. Both visualization methods compute gradients for a given image category related to feature maps of the last convolutional layer to create a class-discriminative region. In our experiment, we used a total of 2,905 chest X-ray images, comprising three cases (such as COVID-19, normal, and viral pneumonia), and empirical evaluations revealed that the proposed method extracted more significant features expeditiously related to the suspected disease. The experimental results demonstrate that our proposed method significantly improves performance metrics: accuracy, precision, recall, and F1 scores reach $96.58\%$, $96.58\%$, $96.59\%$ and $96.58\%$, respectively, which is comparable or enhanced compared with the state-of-the-art methods. We believe that our contribution can support resistance to COVID-19, and will adopt for COVID-19 screening in AI-based systems.
\end{abstract}

\keywords{COVID-19 \and Chest X-ray \and Convolutional Neural Network \and Artificial Intelligence \and Parallel dilation \and Dilation rate  }

\section{Introduction}
Coronavirus or covid-19 is a contagious disease that was caused by the SARS-CoV-2 (Severe Acute Respiratory Syndrome Coronavirus 2). The disease was first discovered and became prevalent in Wuhan, Hubei Province, China, and has since spread around the world. As we know, on March 11, 2020, the World Health Association (WHO) proclaimed the flare-up of coronavirus pandemic \cite{whowebsitesituationreports}. Until now, coronavirus pandemic has devastatingly affected the prosperity of the world economy and health system. As of July 12, 2020, more than 12,401,262 confirmed cases of COVID-19 and 559,047 confirmed deaths due to the disease \cite{whowebsite}. Contemporaneous observations detail that COVID-19 infection causes lung illness, hence, making it troublesome to breathe. As well, a person infected with coronavirus spread the virus among others primarily through coughs or sneezes as tiny droplets, therefore, a legitimate social framework for a disengagement is obligatory to delimit the disease. While attempting to deal with growing numbers of COVID-19 infected cases, all nations try to increase testing to reduce the pandemic ahead of time. At present, it is vital to extend an effective screening strategy within reach to distinguish COVID-19 cases and segregate infected individuals from others while hospitals are striving with scaling up abilities to meet the increasing number of patients \cite{2020arXiv200309871W}. An ultimate goal against COVID-19 battle is to provide prompt treatment to infected individuals using a viable screening technique. As we all know, reverse transcriptase-polymerase chain reaction (RT-PCR) is a reliable and well-documented screening technology, which has been accepted by researchers and clinical staff, but the flexibility and strict necessity of conditions at the clinical laboratory will greatly delay the accurate detection of suspicious patients \cite{unknown}. Moreover, starting the results of RT-PCR testing in China indicated moderately poor sensitivity, and this issue occurred due to the variable characteristics of testing \cite{2020arXiv200309871W}. Further, subsequent revelations illustrated a noteworthy factor that positive rate declined with time after symptom incipience at once relied on how the specimen collected \cite{Estimatingfalse-negative}.

On the other hand, the limitations of RT-PCR testing have prompted researchers to find a rapid and definitive method for diagnosing COVID-19 infection. Even though COVID-19 pandemic will significantly affect the global healthcare system, a radiography testing such as a chest X-ray would be a discretionary screening technique that has as well utilized for COVID-19 diagnosis. Certainly, chest X-ray is beneficial in emergency diagnosis and treatment considering, the way that this system is quick and simple to operate, and radiologists can yet recognize. In the prior observations, researchers observed that patients exhibit inconsistencies in chest X-ray images that are typical for those infected with COVID-19 \cite{articleRadiology}. However, a few British hospitals are presently turning out artificial intelligence (AI) based disease recognition systems to look for potential symptoms of Covid-19 infection in chest X-ray images, such as patterns of opacity in the lungs \cite{bbcreports}. Significant diagnostic accuracy in medical imaging, which is an indispensable objective of automatic disease detection, has already achieved in recent years using CNN, as the core advancement of the rising AI. Inspired by this, recently, a lot of research has already accomplished by utilizing CNN modalities to detect COVID-19 using chest X-ray images, and has obtained promising results \cite{2020arXiv200309871W,narin2020automatic,2020arXiv200408379R,Apostolopoulos_2020,karim2020deepcovidexplainer}.

The purpose of this research is to extend the curative effect using chest X-ray images showing COVID-19 infection. In this regard, we consider the CNN-based framework as it is renowned for its excellent recognition performance in object classification. In CNN, convolution is one of the core operations to regulate the architecture through its dilation rate. The traditional discrete convolution performs the convolution process at a dilation rate of 1, while the dilated convolution \cite{Yu:2016:MCA} uses a different dilation rate to perform the convolution process, and the proposed PDCOVIDNet varies dilation rate in a parallel stack of convolution layer in CNN, thus reflecting more distinctive features. In our experimental evaluation, 219 chest X-ray images of COVID-19, 1341 normal, and 1345 viral pneumonia were collected and processed and used to train, validate, and test the proposed method. The benchmark data set is publicly accessible \cite{chowdhury2020ai}, and the authors of this benchmark generated data from three different open access data repositories containing chest X-ray images \cite{SIRMreports,joseph2020ai,Mooneyreports}. The pipeline of PDCOVIDNet starts with the data augmentation strategy, and then optimizes and fine-tunes the settings to train CNN modalities that expand in parallel, through generating dominant features at different scale shifting receptive fields. Next, the generated features are fused into a neural network system to produce the final prediction. Also, we use two gradient-weighted class activation maps (such as Grad-CAM \cite{8237336} and Grad-CAM++ \cite{8354201}) to aid our system. These maps provide predictive explanations, and can identify important features related to COVID-19 infection. From experimental evaluation, it shows that the proposed method can identify important features related to COVID-19 disease, and the best accuracy achieved is $96.58\%$. The contributions of this paper are as follows:
\begin{itemize}
  \item We propose and develop a novel CNN framework called PDCOVIDNet to detect COVID-19 from chest X-ray images. Our proposed framework uses a dilated convolution in the parallel stack of convolutional blocks that can capture and propagate important features in parallel over the network which enhances detection accuracy significantly.
  \item We visualize the X-ray images to analyze the COVID and non-COVID cases, and further investigate the incorrect classification.
  \item Finally, we empirically evaluate our approach with the state-of-the-art approaches to highlight the effectiveness of PDCOVIDNet in detecting COVID-19.
\end{itemize}

We organize the rest of this paper in the following ways. Section \ref{Related Works} reviews the state-of-the-art models used in detecting COVID-19 using chest X-ray images. The benchmark dataset and the augmentation strategy are described in Section \ref{benchmark}. Next, Section \ref{materialsandmethodology} explains the main details of the proposed model and its adjustment to the detection of COVID-19 cases. In Section \ref{Experimental Results}, we provide the experimental results and show the comparison between PDCOVIDNet and other models. Observation on visualization techniques and incorrect classification results are illustrated in Section \ref{Visualization using Grad-CAM and Grad-CAM++} and Section \ref{Investigation on the incorrect classification}, respectively. Finally, Section \ref{Conclusion} provides the conclusion of this paper with the future research direction.

\section{Related Works}
\label{Related Works}

With the rapid spread of COVID-19 in many countries around the world, imaging technology can quickly detect COVID-19, which helps to control the spread of disease. Chest X-ray is a promising imaging technology with a historical prospect of an image diagnosis system. It can be fully explored through various feature extraction methods, thereby playing an important role in the diagnosis of COVID-19 disease.

Due to the need for a faster interpretation of chest X-ray images, a CNN-based AI system provides \cite{bullock2020mapping} a comprehensive overview of the latest application areas of AI for COVID-19, which mentions medical imaging for diagnosis. Halgurd et al. \cite{maghdid2020diagnosing}  proposed a customized CNN and pre-trained AlexNet \cite{10.5555/2999134.2999257} model through chest X-ray images, the results showed very promising accuracy in detecting patients infected either COVID-19 or normal with 98\% and 94.1\% accuracy respectively. By using the pre-trained ResNet50 \cite{inproceedings} model, Narin et al. \cite{narin2020automatic} obtained 98\% accuracy. The authors also evaluated several CNN architectures for COVID-19 detection. The Inception \cite{inproceedingsinception} transfer learning model modified to extract radiological features for accurate COVID-19 diagnosis, and then experienced internal and external validation with an accuracy of 89.5\% \cite{ShuaiWang}. Also, the CNN architecture based on transfer learning, named as COVID-Net\cite{2020arXiv200309871W}, used to classify chest X-ray images into three categories: normal, non-COVID and COVID-19 infections. It showed that COVID-Net achieved the best accuracy by achieving a test accuracy of 93.3\%. In \cite{Apostolopoulos_2020}, transfer learning adopted in state-of-the-art convolutional neural network architectures, results demonstrated the proof-of-principle for using CNN with transfer learning to extract radiological features from small data sets. Biraja et al.\cite{ghoshal2020estimating} investigated the uncertainty in the CNN solution based on Bayesian Convolutional Neural Network (BCNN) to improve the diagnostic performance of COVID-19. Another CNN based on ResNet50 used by Bukhari et al.\cite{diagnosticevaluation} to prove the usefulness and diagnostic accuracy of patients infected with COVID-19  having accuracy of 98.18\%. The authors\cite{minaee2020deepcovid} of proposed a deep learning architecture for COVID-19 detection utilizing their benchmark in which images exhibit COVID-19 disease verified by radiologists, by fine-tuning four renowned pre-trained convolutional models (ResNet18\cite{inproceedings}, ResNet50, SqueezeNet \cite{i2016squeezenet}, and DenseNet121 \cite{inproceedingsdense}), thereby yielded promising results in terms of sensitivity and specificity. A deep learning framework based on EfficientNet\cite{tan2019efficientnet} proposed by Eduardo et al. \cite{luz2020effective}. Contrasted with some popular baseline models (such as VGG16\cite {simonyan2014deep} and ResNet50), the learning parameters have greatly reduced, and the COVID-19 case has a 100\% positive prediction. Farooq et al. \cite{farooq2020covidresnet} proposed a method called COVID-ResNet that uses a three-step technique, which includes gradually adjusting the size, looping the learning rate search and discriminating the learning rate, and then fine-tuning the pre-trained ResNet50 architecture to improve model performance. A deep learning model called DarkCovidNet\cite {Ozturkdeep} proposed for the automatic detection of COVID-19 using chest X-ray images. The average classification accuracy of binary classification (such as COVID and No-Findings) was 98.08\%, and the average classification accuracy of multi-class classification (such as COVID, No-Findings and pneumonia) was 87.02\%. Finally, the author provided an intuitive explanation and evaluated by expert radiologists.

In a case of AI-based COVID-19 diagnostics system, it is obligatory to verify results with expert radiologists. Almost all methods have shown encouraging results. To the best of our knowledge, a few methods are different only if they have verified by radiologists. Several methods focus only on quantitative analysis, while other methods focus on qualitative analysis using visualization and localization techniques to prove that their analysis can be used for COVID-19 detection, which aids to allow for human-interpretable explanation. In literature reviews, there are a few model and prediction level integration strategies to mitigate variance errors and enhance the performance and generalization of CNN models. Finally, due to the inadequate number of COVID-19 cases, creating ample benchmarks is a major challenge in COVID-19 detection. Considering a small data set, running a large number of iterative CNN architectures can lead to overfitting.

\section{Data Pre-processing}
\label{benchmark}

First, we will introduce the benchmark data set and the expansion strategy in detail to aid the training of the proposed model. Next, we will discuss in detail the proposed PDCOVIDNet architecture design method and the training strategy covering the optimal parameter adjustment. Finally, in order to make suspicious disease detection more convincing, we will integrate visualization techniques to highlight key issues with visual markers.

\subsection{Chest X-ray image benchmark}

\begin{figure}[!ht]
  \centering
  \begin{subfigure}[b]{0.3\linewidth}
    \includegraphics[width=\linewidth]{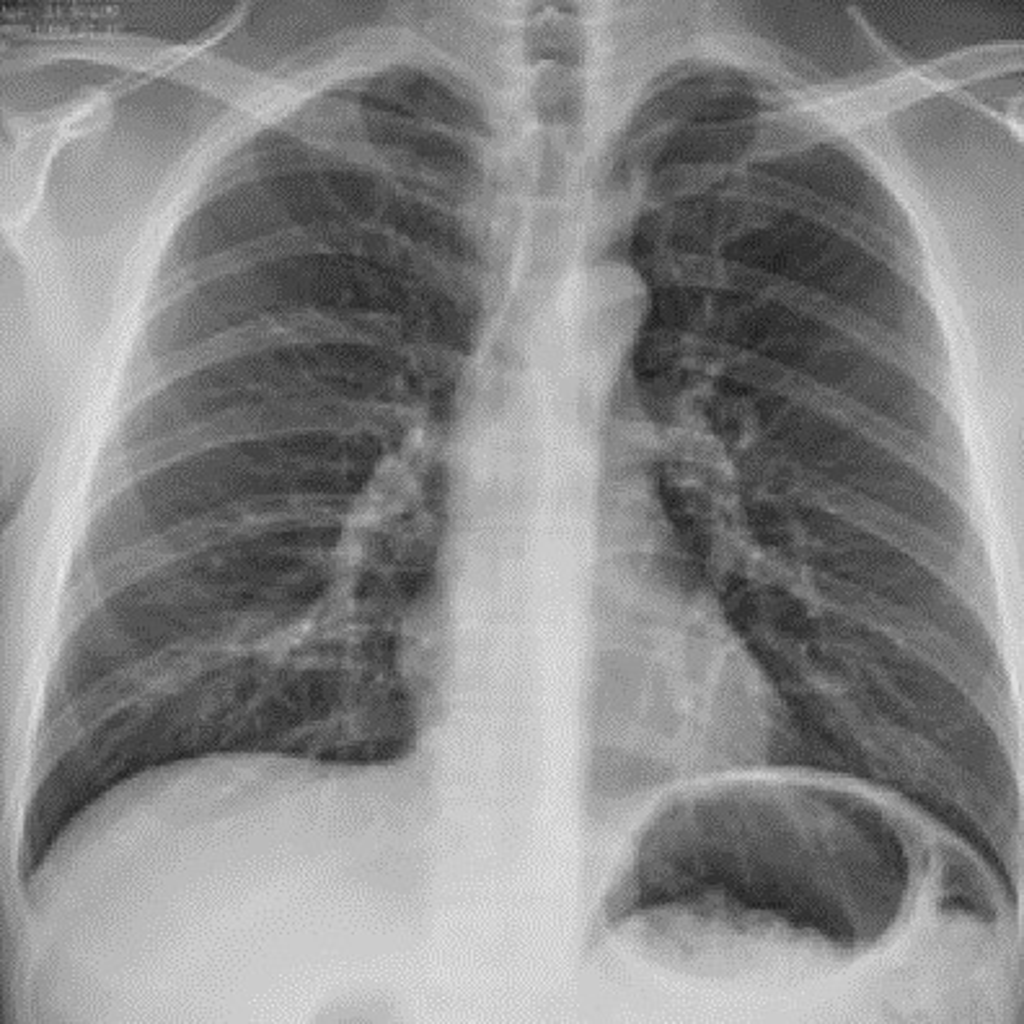}
    \captionsetup{justification=centering}
    \caption{COVID-19}
  \end{subfigure}
  \begin{subfigure}[b]{0.3\linewidth}
    \includegraphics[width=\linewidth]{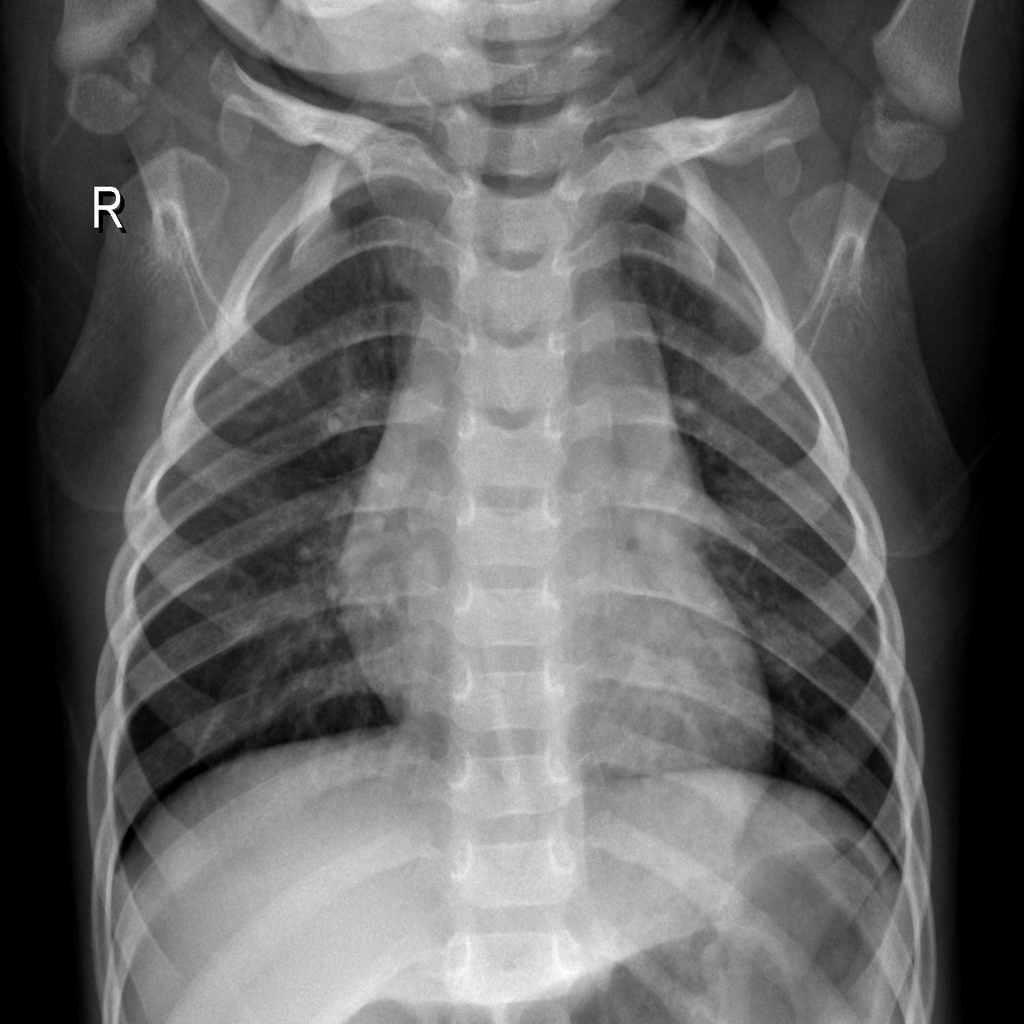}
    \captionsetup{justification=centering}
    \caption{Normal}
  \end{subfigure}
  \begin{subfigure}[b]{0.3\linewidth}
    \includegraphics[width=\linewidth]{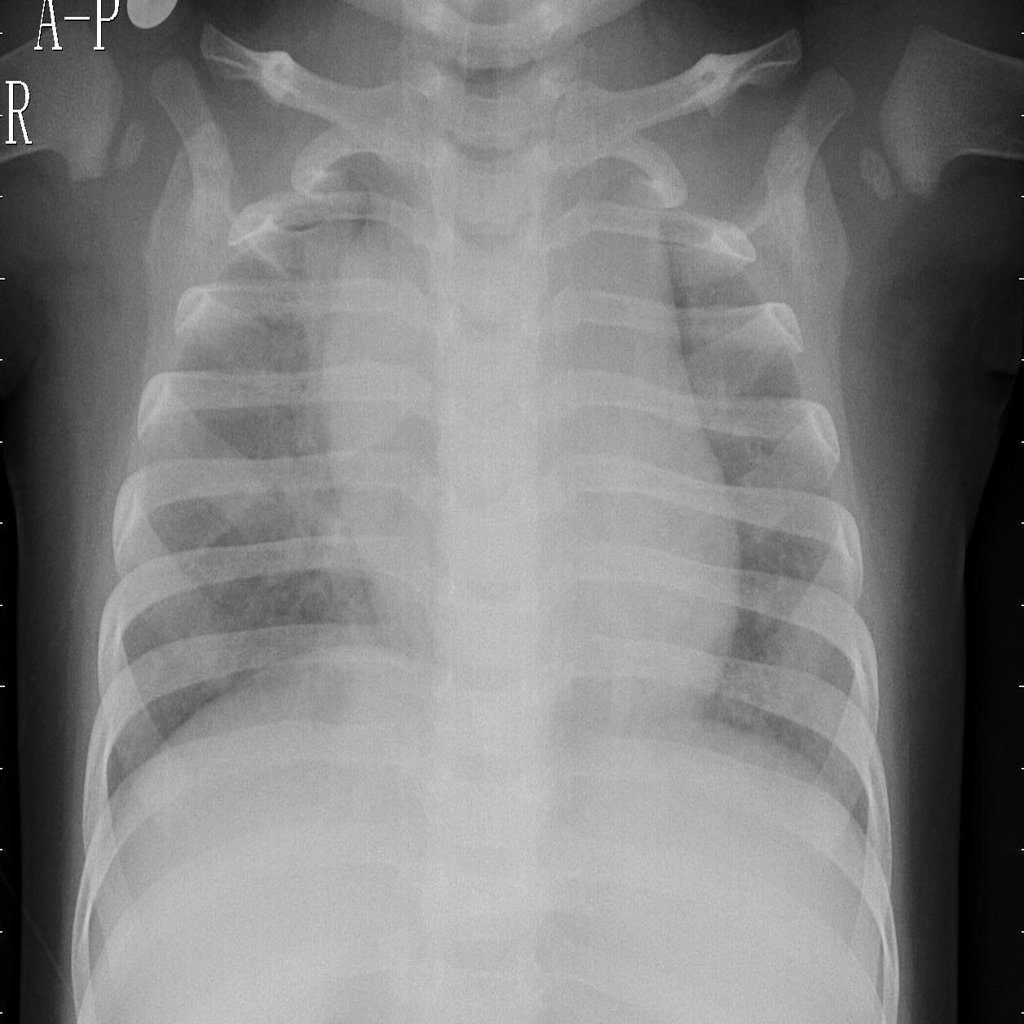}
    \captionsetup{justification=centering}
    \caption{Viral Pneumonia}
  \end{subfigure}
  \captionsetup{justification=centering}
  \caption{Some image labels available in the benchmark dataset \cite{chowdhury2020ai}.}
  \label{fig:benchmark}
\end{figure}

The benchmark data set\cite{chowdhury2020ai} used in our experimental evaluation consists of three main categories (such as COVID-19, Normal, and Viral Pneumonia), yielding 219 COVID-19 positive, 1341 normal and 1345 viral pneumonia chest X-ray images. In the case of accumulating COVID-19 positive images, the authors used two open-access repositories, such as the Italian Society of Medical and Interventional Radiology (SIRM) COVID-19 database\cite{SIRMreports}, and the Novel Corona Virus 2019 dataset developed by Joseph et al.\cite{joseph2020ai}. On the other hand, P. Mooney\cite{Mooneyreports} created normal and viral pneumonia images from the chest X-ray image (pneumonia) database used for this benchmark\cite{chowdhury2020ai}. Moreover, the benchmark is public, and metadata is distributed to provide appropriate document guidance to generate references to each image. Since resizing is one of the essential steps in data preprocessing, all images resized to $224\times 224$ pixels. Figure \ref{fig:benchmark} shows sample images from the benchmark dataset, including COVID-19, normal and viral pneumonia. As shown in the Table \ref {image partition}, we have trained, validated and tested the images in an appropriate ratio.

\begin{table}[!ht]
\centering
\caption{Images partition in Training, Validation and Testing}
      \begin{tabular}{ccccc}
        \hline
        Category  & COVID-19  & Normal   & Viral Pneumonia & Total\\ \hline \hline
        Training   & 175        & 1072        & 1076                 & 2323 \\
        Validation & 21        & 134        & 134                 & 289 \\
        Testing & 23        & 135        & 135                 & 293 \\ \hline
      \end{tabular}
      \label{image partition}
\end{table}

\subsection{Data Augmentation}

In order to properly train the CNN model, it is often useful to manually increase the size of the data set using data enhancement that reduces noise and preserves the original quality. This process is executed just-in-time during the training process, so the performance of the model can be improved by solving the problem of overfitting. For image augmentation, we have many options to choose values of different scales, including horizontal flip, height and width offset, rotation, shearing, zoom, and fill modes. Each option has its ability to represent images in different ways to provide important features during the training phase and thus enhances the model’s performance better. Table \ref{augmentation settings} shows the image augmentation settings used in our experiment.

\begin{table}[!ht]
\centering
\caption{Images augmentation settings}
      \begin{tabular}{cc}
        \hline
        Option  & Value \\ \hline \hline
        rotation range     & 30       \\
        height shift  & 0.15   \\
        width shift  & 0.15    \\
        shear range    & 0.10   \\
        zoom range & 0.10    \\
        fill mode      & nearest      \\ \hline
      \end{tabular}
      \label{augmentation settings}
\end{table}

\section{PDCOVIDNet Architecture}
\label{materialsandmethodology}

In this section, we will briefly describe our proposed PDCOVIDNet architecture. In our proposed model, we have three main components, such as feature extraction, detection, and visualization. First of all, our proposed PDCOVIDNet is a parallel stack of convolutional layers, activation layer and max-Pooling layer. Then, we add parallel layers at the feature level, and perform a convolution again with activation on resulting feature maps. Afterward, the flattened features provide into two layers of Multi-Layer Perceptrons (MLP), but an adjustment needs to determine the proportion of neurons at each layer that drop, to avoid overfitting. Finally, the last layer with softmax activation function performs the classification task, and then generates a class activation map, which acts as an interpreter of classification, merged with the last convolution layer. Figure \ref {fig:PDCOVIDNet System} shows the overall system architecture of the proposed PDCOVIDNet. We split the workflow into two parts: the feature extraction phase and the classification and visualization phase. In the next section, we will briefly explain the feature extraction process.

\subsection{Feature Extraction}

\begin{figure}[!ht]
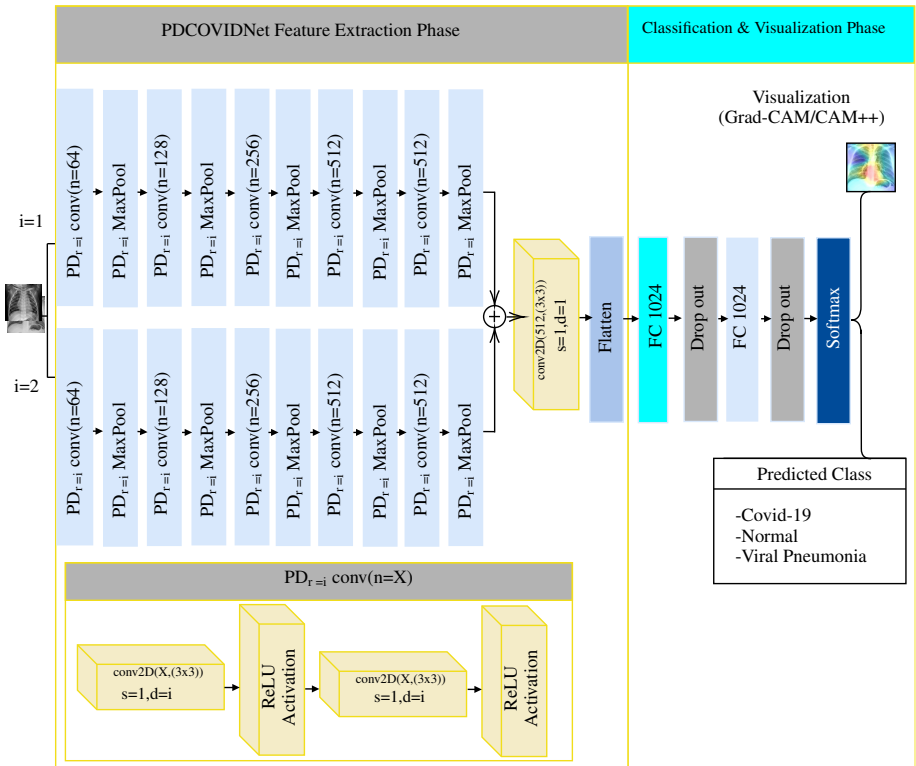

    \resizebox{12.4cm}{!}{

\tikzset{every picture/.style={line width=0.75pt}} 


    }
    \caption{An overall system architecture of PDCOVIDNet.}
    \label{fig:PDCOVIDNet System}
        \end{figure}

\begin{figure}[!ht]
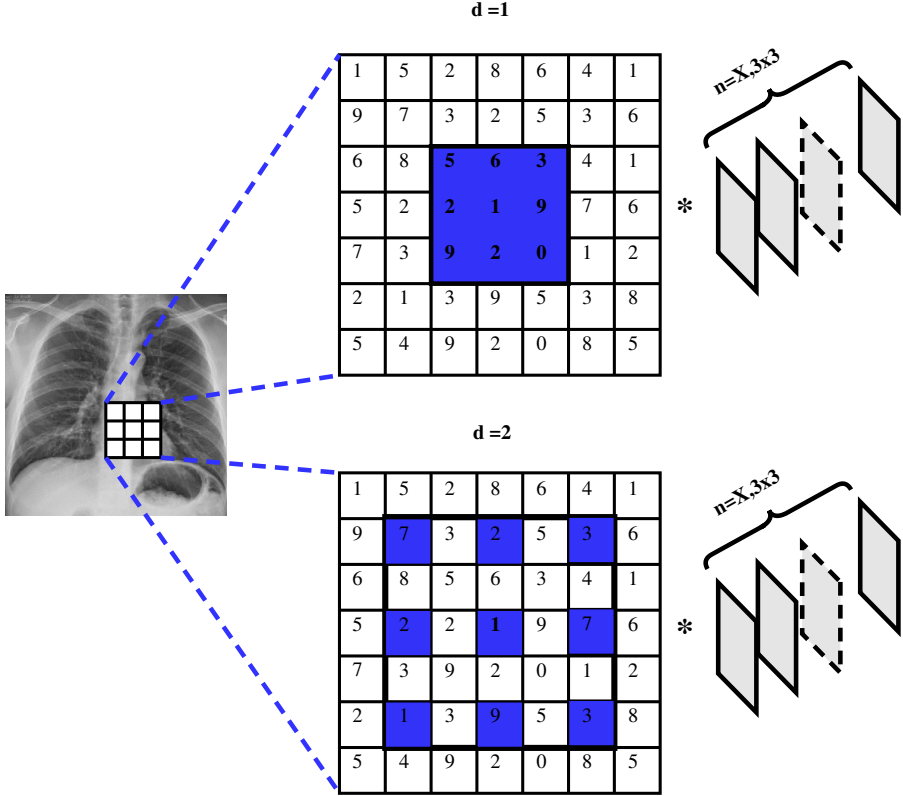

    \resizebox{12.2cm}{!}{
    
    \tikzset{every picture/.style={line width=0.75pt}} 
    
    %
    }
    \caption{Dilated convolution and its activation in PDCOVIDNet. At the top, we have a dilated convolution with a dilation rate of $1$, and the corresponding receive field size is $3 \times 3$, which is equivalent to the standard convolution. It has a $5 \times 5$ receiving field size at the bottom and a dilation rate of $2$. All dilated convolutions have a kernel size of $3 \times 3$ and n filters.}
    \label{fig:dilatedCNN}
        \end{figure}

To obtain a suitable network architecture, different numbers of filters in each convolution layer, filter sizes, different layers in MLP and different hyper-parameters have experimented. In the first stage, PDCOVIDNet consists of five dilated convolutional blocks, expressed as PD\textsubscript{r=i}conv(n=X), that are alternately max-pooled. Figure \ref{fig:dilatedCNN} illustrates how dilated convolution is incorporated into our proposed model. As shown in Figure \ref{fig:dilatedCNN}, the input image provides in two PD\textsubscript{r=i}conv(n=X) blocks in parallel, only changing the dilation rate, such as $d=1,...,N$. A convolution with a dilation rate of $1$ is equivalent to a standard convolution, while a convolution with a dilation rate of greater than $1$ expands the receptive field when processing input at a higher resolution, thereby achieving fine details of the image. The receptive field refers to the portion of the image where a filter extracts feature without change filter size, and is simply an input with a fixed gap, i.e., if there is dilation of $d$, then each input skips of $d-1$ pixels. According to this definition, considering that our input is a $2D$ image, the dilation rate of $1$ is a standard convolution, and dilation rate of $2$ means that each input skips a pixel. To understand the relation between the dilation rate of $d$ and the receptive field size of $rf$, it is often useful to understand the effect of $d$ on $rf$ when the kernel is fixed in size.  Equation \ref{receptivefield}\cite{dumoulin2016guide} depicts the form of a receptive field size where the kernel of size $k$ is dilated by the factor $d$.

\begin{equation}
\begin{aligned}
{rf} &= d(k-1)+1.
\label{receptivefield}
\end{aligned}
\end{equation}
Equation \ref{receptivefield} refers to form the following equation that renders the size of the output $o$, where $m\times m$ input with a dilation factor, padding and stride of $d$, $p$ and $s$ respectively.
\begin{equation}
\begin{aligned}
o &= \left\lfloor\dfrac{m+2p-rf}{s}\right\rfloor+1.
\label{receptiveimageoutput}
\end{aligned}
\end{equation}

After using two receptive fields of different sizes, it captures important features in the observation area at different scales. In our proposed model, each PD\textsubscript{r=i}conv(n=X) block includes two convolution layers and activation layers, where each block consists of a different number of filter response (i.e. X=[$64,128,256,512,512$]) of $3 \times 3$ filter size having stride and dilation rate with $1$ and $d$, respectively.  We define a convolution layer with filters $\mathbf{F} \in \mathbb{R}^{1\times n}$ given as
\def\A{
\begin{bmatrix}
    f_1^{k\times k} & f_2^{k\times k} & \cdots & f_{n=X}^{k\times k}
\end{bmatrix}}
\begin{equation}
\mathbf{F} =\A,
\end{equation}
where $k\times k$ is filter size and $X$ is total number of filters. For a dilated convolution with a dilation rate of $d$ on $m\times m$ input feature maps at layer $l$ , the convolution generates feature maps from the input, denoted as $\mathbf{X}_l^{m\times m}$, and calculated by
\begin{equation}
\mathbf{X}_l^{o\times o} =\mathbf{X}_{{l-1},{d=i}}^{m\times m}*\mathbf{{F}}_l+\mathbf{{B}}_l,
\end{equation}
where $\mathbf{{B}}_l$ is a bias of the $l$-layer, and $\mathbf{{F}}_l$ is the $l$-layer filter of size $k \times k$. The features of layer $l$ are generated at the dilation rate of $d$ with the feature map in layer $(l-1)$. After the convolution layer, we introduce a nonlinear layer with an activation function that uses the features generated at an earlier stage to create a new feature map as output. In the case of activation, we prefer the rectified linear unit (ReLU)\cite{conf/icml/NairH10} because it can integrate the nonlinear layer and the rectification layer in CNN. ReLU has several advantages, and most importantly, it can effectively propagate gradients. Therefore, if the initial weight takes into account the unique characteristics of CNN, the possibility of gradient disappearance can be reduced. Note that the activation function performs element-by-element operations on the input feature map, so the output is the same size as the input. Assuming that the layer $l$ is the active layer of the $n$-th filter, it obtains the input feature $\mathbf{X}_{{(l-1)},{n}}^{m\times m}$ with the feature map $m\times m$ from the previous convolution layer, and generates the same number of features defined as:
\begin{equation}
\mathbf{X}_{l,n}^{m\times m} =max(\mathbf{X}_{{(l-1)},{n}}^{m\times m},0)
\end{equation}
where $\mathbf{X}_{l,n}^{m\times m}$ maps negative values to zero.

The proposed model has a PD\textsubscript{r=i}conv(n=X) block, followed by a max-Pooling layer, and is set five times in parallel successive form. The proposed model has a PD\textsubscript{r=i}conv(n=X) block, followed by a max-Pooling layer five times in parallel successive form. Max-Pooling, which takes the maximum value in each window, is an efficient approach to downscale the filtered image, because when using a filter size of $2\times 2$ with a stride of $2$, three-fourths of generated features are ignored in each layer substantially it reduces the computational complexity for the next layer. The max-pooling window used in our experiment was $2\times 2$ and the stride was $2$, because as reported in earlier studies\cite{inproceedingspooling}, the overlapping window did not improve significantly over the non-overlapping window. Then, in Figure \ref{fig:PDCOVIDNet System}, we see that the features generated from the parallel branches are concatenated and provided to the next convolution layer. The inspiration behind this concatenation-convolution operation is that each branch generates features from images at different layers of CNN have different properties, so we concatenate low-level features of parallel branches to explore feature relationship of dilated convolution hence final convolution layer might detect dominant features for classification. In the last convolution layer, a total of $512$ filters with a filter size of $k\times k$ and a dilation rate of $1$ are applied to create final low-level features followed by ReLU activation. After that, we inaugurate a flatten layer, which converts the square feature map into a one-dimensional feature vector and prepares it for the next phase, which is finally a classification task. Our final task is the classification and visualization phase, which will briefly illustrate in the next section.

\subsection{Classification and Visualization}

At this stage, a two-layer MLP (often called a fully connected (FC) layer) feeds the results of the flatten layer through two neural layers to perform the classification task. It attempts to render the activation from the previous FC layers into class scores (i.e., in classification). In addition, we include a Dropout \ cite {JMLR: v15: srivastava14a} layer after each FC layer. This layer can randomly discard some FC layer weights during training to reduce overfitting. The number of randomly selected weight drops is defined by the dropout limit, which ranges from $0$ to $100\%$.  Indeed, the best adjustment is to determine the optimal number of weights to use in each layer and the dropout ratio to avoid overfitting at the same time, making the network more robust. In this study, we chose a dropout size of $0.3$, two FC layers of size $1024$ and $1024$, respectively, and used the softmax activation function to determine the classes of the input chest X-ray images as COVID-19, normal and viral pneumonia. Finally, the layer details of the proposed model are shown in Table \ref{PDCOVIDNet Parameters}.
\begin{table}[!ht]
\centering
\caption{PDCOVIDNet layer details}
      \begin{tabular}{ccccc}
        \hline
        Layer  & filter size/X, Stride  & PDCOVIDNet (Output Size)\\ \hline \hline
        Input   & -     & \begin{tabular}{c}Branch \\\hline 
        \begin{tabular}{cc}PD\textsubscript{r=1}conv(n=X) & PD\textsubscript{r=2}conv(n=X) \end{tabular}\\ \end{tabular}\\ \hline \hline
        
        Conv2D & $3 \times 3/64,1$        & \begin{tabular}{cc}$224 \times 224 \times 64$ & $224 \times 224 \times 64$ \end{tabular}\\ \hline 
        Conv2D & $3 \times 3/64,1$        & \begin{tabular}{cc}$224 \times 224 \times 64$ & $224 \times 224 \times 64$ \end{tabular} \\ \hline
        MaxPooling2D & $2 \times 2/64,2$        & \begin{tabular}{cc}$112 \times 112 \times 64$ & $112 \times 112 \times 64$ \end{tabular} \\ \hline
        
        Conv2D & $3 \times 3/128,1$        & \begin{tabular}{cc}$112 \times 112 \times 128$ & $112 \times 112 \times 128$ \end{tabular}\\ \hline 
        Conv2D & $3 \times 3/128,1$        & \begin{tabular}{cc}$112 \times 112 \times 128$ & $112 \times 112 \times 128$ \end{tabular} \\ \hline
        MaxPooling2D & $2 \times 2/128,2$        & \begin{tabular}{cc}$56 \times 56 \times 128$ & $56 \times 56 \times 128$ \end{tabular} \\ \hline
        
        Conv2D & $3 \times 3/256,1$        & \begin{tabular}{cc}$56 \times 56 \times 256$ & $56 \times 56 \times 256$ \end{tabular}\\ \hline 
        Conv2D & $3 \times 3/256,1$        & \begin{tabular}{cc}$56 \times 56 \times 256$ & $56 \times 56 \times 256$ \end{tabular} \\ \hline
        MaxPooling2D & $2 \times 2/256,2$        & \begin{tabular}{cc}$28 \times 28 \times 256$ & $28 \times 28 \times 256$ \end{tabular} \\ \hline
        
        Conv2D & $3 \times 3/512,1$        & \begin{tabular}{cc}$28 \times 28 \times 512$ & $28 \times 28 \times 512$ \end{tabular}\\ \hline 
        Conv2D & $3 \times 3/512,1$        & \begin{tabular}{cc}$28 \times 28 \times 512$ & $28 \times 28 \times 512$ \end{tabular} \\ \hline
        MaxPooling2D & $2 \times 2/512,2$        & \begin{tabular}{cc}$14 \times 14 \times 512$ & $14 \times 14 \times 512$ \end{tabular} \\ \hline
        
        Conv2D & $3 \times 3/512,1$        & \begin{tabular}{cc}$14 \times 14 \times 512$ & $14 \times 14 \times 512$ \end{tabular}\\ \hline 
        Conv2D & $3 \times 3/512,1$        & \begin{tabular}{cc}$14 \times 14 \times 512$ & $14 \times 14 \times 512$ \end{tabular} \\ \hline
        MaxPooling2D & $2 \times 2/512,2$        & \begin{tabular}{cc}$7 \times 7 \times 512$ & $7 \times 7 \times 512$ \end{tabular} \\ \hline
        
        Add  & -  & $7 \times 7 \times 512$\\ \hline
        Conv2D  & $3 \times 3/512,1$  & $7 \times 7 \times 512$\\ \hline
        Flatten  & -  & $25088$\\ \hline
        FC  & -  & $1024$\\ \hline
        FC  & -  & $1024$\\ \hline
        Softmax  & -  & $3$\\ \hline

      \end{tabular}
      \label{PDCOVIDNet Parameters}
\end{table}

Although CNN models are powerful in producing impressive results, there are still many questions about why and how to produce such good results. Owing to its black-box nature, it is sometimes challenging to adopt it in a real-life application (such as a medical diagnosis system) where we need an interpretable model. However, early studies\cite{8237336,8354201,cvpr2016_zhou} focused their attention on visualizing the behavior of CNN models, and various visualization methods emphasized the importance of distinguishing classes, so they could execute models with interpretability. In our proposed model, we use Grad-CAM and Grad-CAM++ to highlight the important regions that are class-discriminative saliency maps, where the model emphasizes a gradient-based approach that computes the gradients for a target image class on the feature maps of the final convolution layer in a CNN model. For a given image, let $A^{k}_{i,j}$ denotes the activation map at a spatial location ($i,j$) for the $k$-th filter. The class-discriminative saliency map $L^{c}$ for the target image class $c$ is then computed as\cite{8237336}:
\begin{equation}
\begin{aligned}
L^{c}_{i,j} &= ReLU(\sum_{k}w^{c}_{k}A^{k}_{i,j})
\label{classssaliencymap}
\end{aligned}
\end{equation}
In Eq.\ref{classssaliencymap}, the role of ReLU is to capture features that have a positive impact on the target class. Then, in the case of Grad-CAM, gradients that are flowing back to the final convolutional layer are globally averaged to calculate the target class weights of the $k$-th filter, as described in Eq.\ref{weightclassssaliencymapGradCAM}. Here, $Z$ is the total number of pixels in the activation map, and $Y^{c}$ is the probability that the target category is classified as $c$.  
\begin{equation}
\begin{aligned}
w^{c}_{k} &= \frac{1}{Z}\sum_{i}\sum_{j}\frac{\partial Y^{c}}{\partial A^{k}_{i,j}}
\label{weightclassssaliencymapGradCAM}
\end{aligned}
\end{equation}
On the other hand, Grad-CAM++ contributes to the weighted average of pixel-level gradients rather than the global average of gradients, so the pixel weights in a particular feature map contribute to the overall decision of detection. Grad-CAM++ redevelops Eq.\ref{weightclassssaliencymapGradCAM} to ensure that their contribution to the weighted average of the gradients remains unchanged without losing generality, i.e.,
\begin{equation}
\begin{aligned}
w^{c}_{k} &= \sum_{i}\sum_{j}\alpha ^{kc}_{i,j}.ReLU(\frac{\partial Y^{c}}{\partial A^{k}_{i,j}})
\label{weightclassssaliencymapGradCAM++}
\end{aligned}
\end{equation}
\begin{equation}
\begin{aligned}
\alpha ^{kc}_{i,j} &= \frac{\frac{\partial^{2} Y^{c}}{(\partial A^{k}_{i,j})^{2}}}{2\frac{\partial^{2} Y^{c}}{(\partial A^{k}_{i,j})^{2}}+\sum_{a}\sum_{b} A^{k}_{a,b}{\frac{\partial^{3} Y^{c}}{(\partial A^{k}_{i,j})^{3}}} }
\label{weightalphaclassssaliencymapGradCAM++}
\end{aligned}
\end{equation}
In Eq.\ref{weightalphaclassssaliencymapGradCAM++}, $(i, j)$ and $(a, b)$ are iterators over the same activation map $A^{k}$\cite{8354201}.

\section{Experimental Evaluation}
\label{Experimental Results}

In this section, we will present the performance of our proposed model to classify chest X-ray images, broadly categorized into three classes: COVID-19, Normal and Viral Pneumonia. In Section \ref{materialsandmethodology}, a brief description of the benchmark data set used in the experiment and the augmentation approach are discussed. The experiment sets the training, validation, and test ratios to $80\%$, $10\%$, and $10\%$, respectively. We compared our proposed PDCOVIDNet with VGG16, ResNet50, InceptionV3\cite{44903} and DenseNet121, and did not use any pre-trained weights (such as ImageNet) since ImageNet weights come from images of general objects, not chest X-ray images. All of our experiments are executed in Keras with the TensorFlow backend.


\subsection{Hyper-parameters Tuning}
Hyper-parameters become critical because they directly control the behavior of the model, so fine-tuned hyper-parameters have a huge impact on the performance of the model. We used the Adam \cite{kingma2014adam} optimizer to train $50$ epochs for each model with a learning rate of $1e-4$, with a batch size of $32$. In addition, we applied the categorical cross-entropy loss function to the training, which measures the loss between the probability of the class predicted from the softmax activation function and the true probability of the category.

\subsection{Performance Evaluation Metrics}
For experimental evaluations, we utilized several evaluation metrics such as Accuracy, Precision, Recall, and F1 score, i.e.,
\begin{equation}
\begin{aligned}
Accuracy &= \frac{TP+TN}{Total\,Samples}
\label{accuracy}
\end{aligned}
\end{equation}
\begin{equation}
\begin{aligned}
Precision &= \frac{TP}{TP+FP}
\label{Precision}
\end{aligned}
\end{equation}
\begin{equation}
\begin{aligned}
Recall &= \frac{TP}{TP+FN}
\label{Recall}
\end{aligned}
\end{equation}
\begin{equation}
\begin{aligned}
F1 &= 2\times\frac{Precision \times Recall }{Precision + Recall }
\label{F1}
\end{aligned}
\end{equation}

where $TP$ stands for true positive, while $TN$, $FP$, and $FN$ stand for true negative, false positive, and false negative, respectively. The $F1$ score may be a more reliable measure because the benchmark dataset is unbalanced, such as COVID-19 with $219$ images and non-COVID with $2686$ images. Subsequently, we used the ROC (Receiver Operating Characteristics) curve to display the results and measured the area under the ROC curve (often called AUC (Area Under the Curve)) to provide information about the effectiveness of the model.

\subsection{Evaluation of individual model}

The overall results are shown in Table \ref{comparisiondifferentmethods} and Table \ref{weightedavegeragecomparisiondifferentmethods}, where Table \ref{comparisiondifferentmethods} describes the class-wise classification results on different evaluation metrics, and Table \ref{weightedavegeragecomparisiondifferentmethods} shows the weighted average results.\textcolor{blue}{In Table \ref {comparisiondifferentmethods}, inside the square brackets are lower and upper boundaries of the 95\% confidence interval (CI). Compared with specific performance indicators, CIs are considered more practical indicators, the latter can only increase the level of statistical significance. Also, CIs indicate that these results reflect the reliability of the problem domain.} From the table \ref {comparisiondifferentmethods}, we can see that almost all models tend to enhance the classification of most categories (such as normal and viral pneumonia) because they have more training weights than the COVID-19 case. For COVID-19, the highest performance belongs to PDCOVIDNet, whose precision, recall, and F1 scores are $95.45\%$, $91.30\%$, and $93.33\%$, respectively. Our model provides consistent results for the precision, recall, and F1 under normal cases, with each performance index being $97.04\%$, and the recall for ResNet50 is $97.78\%$, slightly higher than PDCOVIDNet. Also, the precision and F1 scores of DenseNet121 are $95.52\%$ and $95.17\%$, respectively, which is comparable to PDCOVIDNet. Next, in the case of viral pneumonia, the precision, recall, and F1 score of PDCOVIDNet are $96.32\%$, $97.04\%$, and $96.68\%$, respectively. In particular, for cases of viral pneumonia, ResNet50 is only $1\%$ better precision than PDCOVIDNet.  In the Table \ref {comparisiondifferentmethods}, the accuracy of all evaluation models is summarized, and it can be seen that PDCOVIDNet is superior to other models. At the same time, it is evident that PDCOVIDNet has the ability to resist class imbalances since COVID-19 cases are smaller than normal or viral pneumonia cases. However, the more structured residual blocks of the model, the worse the classification performance (e.g., ResNet50). As shown in Table \ref{weightedavegeragecomparisiondifferentmethods}, considering the weighted average of all performance evaluation indicators, the best results are obtained by using PDCOVIDNet. In the weighted average comparison, PDCOVIDNet's results are much better than other models, which can be explained by the fact that the proposed model can extract feature maps at different scales from chest X-ray images. In particular, compared with PDCOVIDNet, Densenet121 is missing $ 2 \%$ in each evaluation indicator. Although ResNet50 provides the best performance for normal and viral pneumonia, unexpectedly, it fails to achieve the most successful model in the performance measurement. \textcolor{blue}{In the case of PDCOVIDNet, the tight range of CI means higher precision, while the wide range of other models indicates the opposite.}

\begin{table}[!ht]
\centering
\caption{Class-wise classification results of individual model}
\captionsetup{justification=centering}
\begin{tabular}{llllll}
\hline
Method                      & Class          & Precision & Recall & F1   & Accuracy (95\% CI)                \\ \hline \hline
\multirow{3}{*}{PDCOVIDNet} & COVID-19       & 95.45      & 91.30   & 93.33 & \multirow{3}{*}{ \textbf{96.58[94.51,98.67]}} \\ \cline{2-5}
                            & Normal         & 97.04      & 97.04   & \textbf{97.04} &                         \\ \cline{2-5}
                            & Viral Pnemunia & 96.32      & 97.04   & 96.68 &                         \\ \hline
                            
\multirow{3}{*}{VGG16}      & COVID-19       & 90.48      & 82.61   & 86.36 & \multirow{3}{*}{93.86[91.11,96.61]} \\ \cline{2-5}
                            & Normal         & 93.43      & 94.81   & 94.12 &                         \\ \cline{2-5}
                            & Viral Pnemunia & 94.81      & 94.81   & 94.81 &                         \\ \hline 

\multirow{3}{*}{ResNet50}      & COVID-19       & 94.74      & 78.26   & 85.71 & \multirow{3}{*}{92.15[89.07,95.23]} \\ \cline{2-5}
                            & Normal         & 87.42      & \textbf{97.78}   & 92.31 &                         \\ \cline{2-5}
                            & Viral Pnemunia & \textbf{97.56}      & 88.89   & 93.02 &                         \\ \hline 

\multirow{3}{*}{InceptionV3}      & COVID-19       & 83.83      & 86.96   & 85.11 & \multirow{3}{*}{93.51[90.70,96.34]} \\ \cline{2-5}
                            & Normal         & 96.15      & 92.59   & 94.34 &                         \\ \cline{2-5}
                            & Viral Pnemunia & 92.81      & 95.56   & 94.16 &                         \\ \hline 

\multirow{3}{*}{DenseNet121}      & COVID-19       & 95.24      & 86.96   & 90.91 & \multirow{3}{*}{94.54[91.94,97.14]} \\ \cline{2-5}
                            & Normal         & 95.52      & 94.81   & 95.17 &                         \\ \cline{2-5}
                            & Viral Pnemunia & 93.48      & 95.56   & 94.51 &                         \\ \hline  
\end{tabular}
\label{comparisiondifferentmethods}
\end{table}

\begin{table}[!ht]
\centering
\caption{Weighted average results of each model}
\captionsetup{justification=centering}
\begin{tabular}{llll} \hline
Method      & Precision (95\% CI) & Recall (95\% CI) & F1     \\ \hline \hline
PDCOVIDNet  & \textbf{96.58 $\pm$ 2.08}    & \textbf{96.59 $\pm$ 2.05} & \textbf{96.58} \\
VGG16       & 93.84 $\pm$ 2.75    & 93.86 $\pm$ 2.75 & 93.83 \\
ResNet50    & 92.67 $\pm$ 2.98    & 92.15 $\pm$ 3.08 & 92.12 \\
InceptionV3 & 93.60 $\pm$ 2.80    & 93.52 $\pm$ 2.82 & 93.53 \\
DenseNet121 & 94.54 $\pm$ 2.60    & 94.54 $\pm$ 2.60 & 94.53 \\ \hline
\end{tabular}
\label{weightedavegeragecomparisiondifferentmethods}
\end{table}

\begin{figure}[!ht]
\centering
\includegraphics[width=\textwidth,height=150pt]{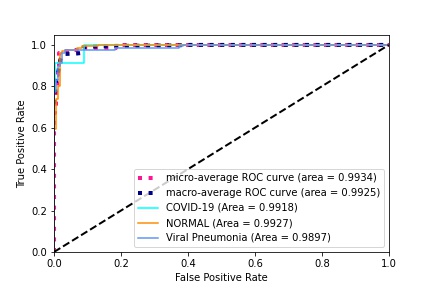}
\caption{Comparison of the ROC curve for COVID-19, Normal and Viral Pneumonia using PDCOVIDNet}
\label{fig:PDCOVIDNet ROC Curve.}
\end{figure}

It is often hard to measure the performance of the model using precision, recall and accuracy, so we need to look at the ROC curve which allows a false positive rate since it plots the true positive rate against a false positive rate. In Figure \ref{fig:PDCOVIDNet ROC Curve.}, ROC curves show the micro and macro average and class-wise AUC scores achieved with the PDCOVIDNet, and show consistent AUC scores across all classes, indicating stable predictions of the proposed model. In ROC curves, we obtained AUC scores of $0.9918$, $0.9927$, and $0.9897$ for COVID-19, normal and viral pneumonia, respectively. We can see that the area under the curve of all classes is relatively similar, but normal's AUC is slightly higher than other classes. \textcolor{blue}{Furthermore, Table \ref{AUC_95_CI} reports the AUC with 95\% CI.}
 
\begin{table}[!ht]
\centering
\caption{Area under the ROC (AUC) and 95\% confidence interval of AUC}
\captionsetup{justification=centering}
\begin{tabular}{ll} \hline
Category      & AUC (95\% CI)     \\ \hline \hline
Micro-Average  & 0.9934 $\pm$ 0.01\\
Macro-Average       & 0.9925 $\pm$ 0.01  \\
COVID-19    & 0.9918 $\pm$ 0.04  \\
Normal & 0.9927 $\pm$ 0.01  \\
Viral Pneumonia & 0.9897 $\pm$ 0.02 \\ \hline
\end{tabular}
\label{AUC_95_CI}
\end{table}

Figure \ref{fig:all confusion matrix} shows the confusion matrices for all evaluated models. In Figure \ref{fig:all confusion matrix}, it is clear that of the 23 test images, two of the COVID-19 images are classified as normal and viral pneumonia, and of the 135 images, only one image of viral pneumonia is related to COVID-19, but none of the normal images belong to COVID-19. One of the reasons may be that COVID-19 is an especial case of viral pneumonia, so they have common features that mislead the PDCOVIDNet model. It becomes clear that the VGG16 model has the same recall for the classification of normal and viral pneumonia, although it shows a significant decline in the positive prediction of COVID-19 cases. However, ResNet50 shows the ability to detect normal images, but in the case of COVID-19 detection, it shows almost the same performance as the VGG16 model, although it shows inadequate performance when predicting viral pneumonia images. As shown in the confusion matrix (Figure \ref{fig:all confusion matrix}(d)), we can say that InceptionV3 can correctly classify more cases of viral pneumonia than COVID-19 and normal cases. Next, DenseNet121 demonstrates the same performance as InceptionV3 in detecting COVID-19, and shows nearly the same performance with VGG16 in detecting viral pneumonia and normal cases. Finally, we can claim that PDCOVIDNet is powerful in detecting COVID-19 cases from chest X-ray images. For this reason, we believe that the proposed model focuses on discriminating features that can help distinguish between other types (e.g., normal and viral pneumonia).

\begin{figure}[!ht]

\vspace{0.11in}

\resizebox{12.2cm}{!}{

\tikzset{every picture/.style={line width=0.75pt}} 



}
\vspace{0.05in}
\caption{Confusion matrix of all evaluated models with test set. Classes 0,1, and 2 represent COVID-19, Normal, and Viral Pneumonia. Lower and upper limits of the 95\% CI are reported inside the square brackets.}
\label{fig:all confusion matrix}

\end{figure}

\section{Visualization using Grad-CAM and Grad-CAM++}
\label{Visualization using Grad-CAM and Grad-CAM++}

\begin{figure}[!ht]

    \resizebox{12.6cm}{!}{
    
    \begin{tabular}{c c c c}
        
        \\
        
        \hspace{1cm}
        &\hspace{1cm}
        &\hspace{1cm}
        &\large True :COVID-19\\
        
        \large Input Image
        &\large Grad-CAM
        &\large Grad-CAM++
        &\large Pred :COVID-19(0.9977$\pm$0.01)\\
        
        \includegraphics[width=3.7cm]{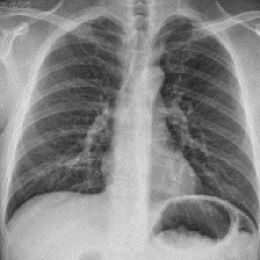} 
        &\includegraphics[width=3.7cm]{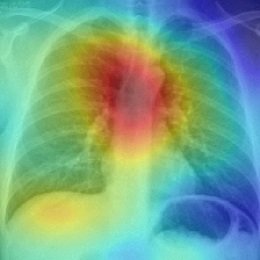}
        &\includegraphics[width=3.7cm]{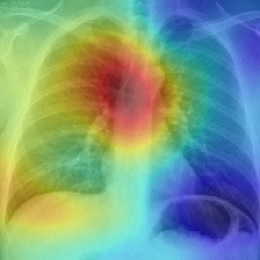}
        &\includegraphics[width=3.7cm]{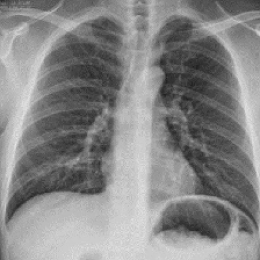}\\\\

        \hspace{1cm}
        &\hspace{1cm}
        &\hspace{1cm}
        &\large True :Normal\\
        
        \large Input Image
        &\large Grad-CAM
        &\large Grad-CAM++
        &\large Pred :Normal(0.8203$\pm$0.04)\\
        
        \includegraphics[width=3.7cm]{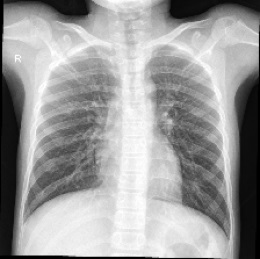} 
        &\includegraphics[width=3.7cm]{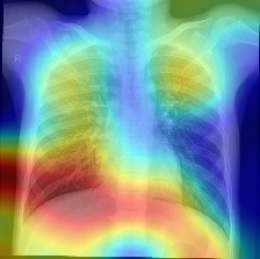}
        &\includegraphics[width=3.7cm]{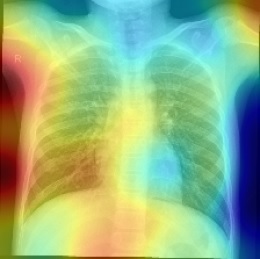}
        &\includegraphics[width=3.7cm]{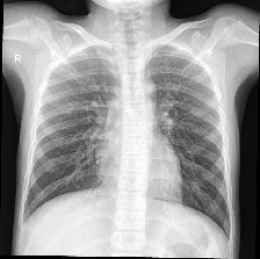}\\\\

        \hspace{1cm}
        &\hspace{1cm}
        &\hspace{1cm}
        &\large True :Viral Pneumonia\\
        
        \large Input Image
        &\large Grad-CAM
        &\large Grad-CAM++
        &\normalsize Pred :Viral Pneumonia(0.9807$\pm$0.02)\\
        
        \includegraphics[width=3.7cm]{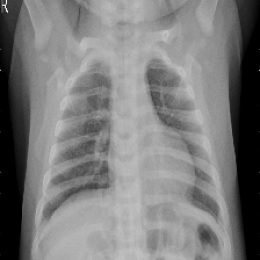} 
        &\includegraphics[width=3.7cm]{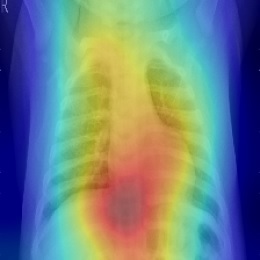}
        &\includegraphics[width=3.7cm]{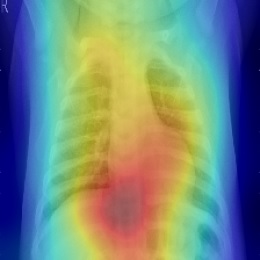}
        &\includegraphics[width=3.7cm]{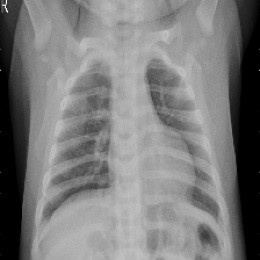}\\\\    
        
        \\
        
    \end{tabular}
    
    }
    
    \caption{Input images and their Grad-CAM, Grad-CAM++, and human-understandable prediction with probability score according to the PDCOVIDNet. The first row provides input with COVID-19, as well as two types of visualization effects, and finally the prediction results with 95\% CI.The second and third rows show the investigation of normal and viral pneumonia, respectively. Here, True means the actual class of the image, and Pred. means the predicted class.}
    \label{fig:PDCOVIDNet Grad-CAM.}

\end{figure}

In our evaluation, we used Grad-CAM and Grad-CAM++ visualization methods to visually represent the salient regions where PDCOVIDNet insisted on making the final classification decision on chest X-ray images. Accurate and decisive salient region detection is important for the interpretation of classification, while also ensuring the reliability of the results. In this regard, a two-dimensional heat map is generated from feature weights with different brightness, which corresponds to the importance of the feature. The heat map is overlaid on the input image to locate the salient region. Figure \ref{fig:PDCOVIDNet Grad-CAM.} shows the visualization results of Grad-CAM and Grad-CAM++ using PDCOVIDNet to locate salient regions when the input image is classified as COVID-19 or normal or viral pneumonia, where the regions distinguishing the classes in the lung have been localized. For COVID 19, both Grad-CAM and Grad-CAM++ generate seemingly the same results, so for the detection of critical areas, the overlapping positions of the heat maps can be considered. In the case of viral pneumonia, the salient regions detected using Grad-CAM and Grad-CAM++ are undifferentiated, while under normal class there are differences, and seems to fail to detect the salient regions as the heat map highlights outside X-ray than inside the lung. To help AI-based systems, it is certainly effective to provide the system with some human-understandable numerical measures (such as probability) as shown in Figure \ref{fig:PDCOVIDNet Grad-CAM.}. 

\section{Investigation on the incorrect classification}
\label{Investigation on the incorrect classification}

\begin{figure}[!ht]

    \resizebox{12.5cm}{!}{
    
    \begin{tabular}{c c c}
        
        \\
        
        \normalsize Pred :Normal(0.54$\pm$0.06) 
        &\normalsize Pred :Viral Pneumonia(0.60$\pm$0.06)
        &\normalsize Pred :Viral Pneumonia(0.88$\pm$0.04) \\
        
        \normalsize True :Viral Pneumonia(0.45$\pm$0.06) 
        &\normalsize True :Normal(0.39$\pm$0.06)
        &\normalsize True :Normal(0.11$\pm$0.04) \\
        
        \includegraphics[width=3.7cm]{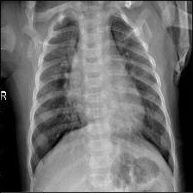} 
        &\includegraphics[width=3.7cm]{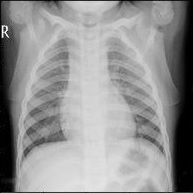}
        &\includegraphics[width=3.7cm]{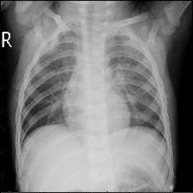}\\\\

        \normalsize Pred :Viral Pneumonia(0.93$\pm$0.03)  
        &\normalsize Pred :Normal(0.93$\pm$0.03) 
        &\normalsize Pred :Viral Pneumonia(0.96$\pm$0.02) \\
        
        \normalsize True :Normal(0.07$\pm$0.03) 
        &\normalsize True :Viral Pneumonia(0.06$\pm$0.03) 
        &\normalsize True :Normal(0.04$\pm$0.02) \\
        
        \includegraphics[width=3.7cm]{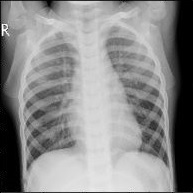}
        & \includegraphics[width=3.7cm]{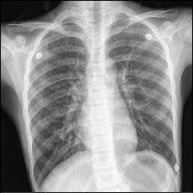}
        & \includegraphics[width=3.7cm]{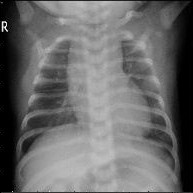}\\\\

        \normalsize Pred :Normal(0.95$\pm$0.02) 
        &\normalsize Pred :Normal(0.98$\pm$0.02)  
        &\normalsize Pred :COVID-19(0.98$\pm$0.02) \\
        
        \normalsize True :COVID-19(0.01$\pm$0.01) 
        &\normalsize True :Viral Pneumonia(0.02$\pm$0.02) 
        &\normalsize True :Viral Pneumonia(0.02$\pm$0.02) \\
        
        \includegraphics[width=3.7cm]{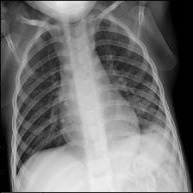}
        & \includegraphics[width=3.7cm]{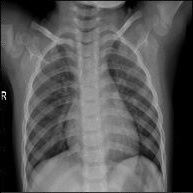}
        & \includegraphics[width=3.7cm]{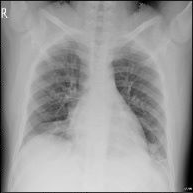}\\\\

        \normalsize Pred :Viral Pneumonia(0.98$\pm$0.02)
        &
        &\\
        
        \normalsize True :COVID-19(0.01$\pm$0.01)
        &
        &\\
        
        \includegraphics[width=3.7cm]{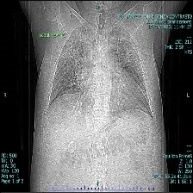}
        &
        &\\\\

    \end{tabular}
    
    }
    
    \caption{Investigation on the incorrectly classified images with the probability of predicted class and actual class. The results for 95\% CI are reported inside the parenthesis.}
    
    \label{fig:PDCOVIDNet Miss Classified Image.}

\end{figure}

In this section, we will further investigate the incorrect classification caused by the use of PDCOVIDNet. The total number of incorrectly classified images is 10, as shown in Figure \ref{fig:PDCOVIDNet Miss Classified Image.}.
Two COVID-19 images are classified as normal and viral pneumonia, and in both cases, COVID-19 is far behind the prediction as the probability of COVID-19 prediction is very low compared with others. Normal images are not classified as COVID-19, but among the four images with normal classification errors, one prediction is on the edge of viral pneumonia, while the other predictions are very different. Correspondingly, among the four incorrectly classified images of viral pneumonia, one image belongs to COVID-19 and the other images are normal, and one of the predictions is very close.

\section{Conclusion and Future Work}
\label{Conclusion}

In this paper, we proposed a CNN-based method, called PDCOVIDNet, for detecting COVID-19 from chest X-ray images. As we have seen, PDCOVIDNet can effectively capture COVID-19 features by dilated convolution in the parallel stack of convolution blocks, so it has an excellent classification performance compared to some well-known CNN architectures. The dataset used in the experiment has a limited number of COVID-19 images, and at once, it is still developing, but data augmentation techniques have able to surmount the challenge as CNN based architecture needs more data for effective training. Our experimental evaluation shows that PDCOVIDNet outperforms the state-of-the-art models, with its precision and recall are $95.45\%$ and $91.3\%$, respectively. As well, PDCOVIDNet demonstrates its potential through other performance metrics such as the weighted average of precision, recall and F1 scores, and finally the overall model accuracy. We apply the proposed model as well as two visualization techniques to identify the class-discriminative regions because they have a greater influence in classifying the input chest X-ray image into its anticipated classes. Finally, we believe that the current findings will hopefully overcome intellectual challenges to detect more cases of COVID-19 and use them to screen for COVID-19 cases in AI-based systems, especially in clinical practice. 

As future work, we will explore and incorporate a diversified data set with more COVID-19 cases to make our proposed model more robust.

\bibliographystyle{unsrt}  
\bibliography{references}  

\end{document}